\documentclass[twoside,reqno]{qts-proc}
\usepackage{epsfig,cite}
\usepackage{amssymb,amsmath}
\usepackage{times}
\begin{document}
\sloppy \raggedbottom
\setcounter{page}{1}

\newpage
\setcounter{figure}{0}
\setcounter{equation}{0}
\setcounter{footnote}{0}
\setcounter{table}{0}
\setcounter{section}{0}

\title{Quantum computing without qubit-qubit interactions}

\runningheads{Almut Beige}{Quantum Computing without Qubit-Qubit Interactions}

\begin{start}

\author{Almut Beige}{1}

\address{The School of Physics and Astronomy, University of Leeds, Leeds LS2 9JT, United Kingdom}{1} 

\begin{Abstract}
Quantum computing tries to exploit entanglement and interference to process information more efficiently than the best known classical solutions. Experiments demonstrating the feasibility of this approach have already been performed. However, finding a really scalable and robust quantum computing architecture remains a challenge for both, experimentalists and theoreticians. In most setups decoherence becomes non-negligible when one tries to perform entangling gate operations using the coherent control of qubit-qubit interactions. However, in this proceedings we show that two-qubit gate operations can be implemented even without qubit-qubit interactions and review a recent quantum computing scheme by Lim {\em et al.} [{\em Phys. Rev. Lett.} {\bf 95}, 030505 (2005)] using only single photon sources (e.g.~atom-cavity systems, NV colour centres or quantum dots) and photon pair measurements.
\end{Abstract}

\end{start}

\section{Introduction}

There are many practical limitations to the implementation of quantum computing. One problem is dissipation, i.e.~the loss of information due to unwanted interactions with the environment. Another one is the general sensitivity of physical processes to parameter fluctuations. For example, if the amplitude of an applied laser field fluctuates by a few percent, this should not result in a failure of the computation. One solution to these problems is to use measurements: They can be used to project a quantum system into any desired state and are commonly used for state preparation in quantum optics experiments.  

However, measurements can also play a much more subtle role in quantum computing. They can provide the main ingredient for the implementation of entangling two-qubit gate operations (see e.g.~Refs.~\cite{nine,distributed2,2,3,10,fifteen,leung} and references therein). Together with single-qubit operations, entangling two-qubit gates are universal for quantum computing. To avoid the destruction of qubits, it is not allowed to measure the qubits directly. Measurements should be performed on ancillas which have interacted and are therefore entangled with the computational qubits \cite{1}. In order to implement a quantum gate, a measurement should be performed on the ancillas in a basis that is mutually unbiased \cite{Wootters} with respect to the computational basis. This ensures that nobody learns anything about the qubits and the relevant information might remain in the computer. 

The most famous example of such a measurement-based quantum computing approach is the linear optics scheme for photonic qubits by Knill, Laflamme and Milburn \cite{2}. However, ancillas and qubits do not have to be of the same physical nature. For example, if the qubits are atoms in a cavity, the ancillas can be the quantised cavity field mode \cite{3}, a common vibrational mode \cite{4,milburn}, or newly generated photons \cite{5,6}. Vice versa, one can use collective atomic states as ancillas for photonic qubits \cite{7,8}. Quantum computing with hybrid systems should help to overcome some of the most pressing problems in existing non-hybrid proposals, including the difficulty of scaling conventional stationary qubit architectures and the lack of practical means for storing single photons in linear optics setups.

In the following, we describe such a hybrid system containing stationary and flying qubits and discuss the idea of Repeat-Until-Success (RUS) quantum computing by Lim {\em et al.} \cite{6}. Each stationary qubit is obtained from two stable ground states of a single photon source. To perform an entangling two-qubit gate operation, photons should be created in each of the respective sources. Afterwards, the photons should pass simultaneously through a linear optics setup, where a two-photon measurement is performed. This measurement results either in the completion of the desired two-qubit gate or induces two correctable single-qubit gates. In the latter case, no quantum information is lost and the gate operation can be {\em repeated until success}.

This paper is organised as follows. In Section \ref{double}, we discuss the basic features of quantum optical photon interference experiments which allow two photon sources to communicate with each other very efficiently. In Section \ref{main} we describe the possible realisation of an eventually deterministic  two-qubit entangling gate between two distant single photon sources. It is shown that RUS quantum computing requires only interference and photon pair measurements in a carefully chosen basis. Finally we summarise our results in Section \ref{conc}.

\section{A two-atom double-slit experiment} \label{double}

\begin{figure}[b]
\centerline{\epsfig{file=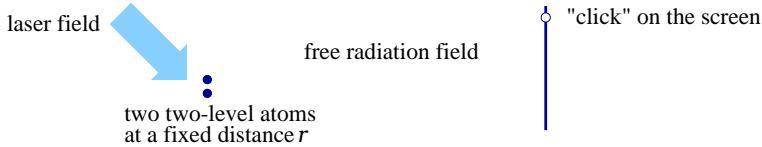,width=100mm}}
\caption{Experimental setup. Two two-level atoms are placed at a fixed distance $r$ from each other. Both are coupled to the same free radiation field and are continuously driven by a resonant laser. This
leads to spontaneous photon emissions. Each photon causes a ``click" at a point on a screen.}\label{doubleslit}
\end{figure}

In 1982, Scully and Dr\"uhl proposed a simple quantum eraser experiment concerning delayed choice phenomena in quantum mechanics \cite{Scully82}. The setup they considered is shown in Figure \ref{doubleslit}. It consists of two two-level atoms trapped at a fixed distance $r$ from each other inside the same free radiation field. The particles are continuously driven by a resonant laser field and spontaneously emit photons. Each emitted photon causes a ``click'' at a certain point on a screen far away from the particles. These ``clicks,''  when collected, add up to an interference pattern with a spatial intensity distribution similar to the one found in classical double-slit experiments. This was verified experimentally by Eichmann {\em et al.} in 1993 \cite{Eichmann}. Since then the interpretation of this experiment attracted continuous interest (see e.g.~Refs.~\cite{Englert,Itano,Schoen} and references therein). 

In this section, we give a short description of the above described two-atom double-slit experiment following the discussion by Sch\"on and Beige \cite{Schoen,schoen2}. They showed that the time evolution of the quantum mechanical components, namely the atoms, the free radiation field and the applied laser light can be modelled with the help of the interaction Hamiltonian
\begin{equation} \label{21}
H_{\rm I} = \hbar \sum_{i=1,2} \sum_{{\bf k}\lambda} {\rm e}^{- {\rm i} (\omega_0-\omega_k) t} \,
{\rm e}^{- {\rm i} {\bf k} {\bf r}_i} \, g_{{\bf k}\lambda}^* \, a_{{\bf k}\lambda}^\dagger \, S_i^- + {\rm H.c.} 
+ H_{\rm laser \, I} \, .
\end{equation}
Here $S_i^- = |1 \rangle_{ii} \langle 2|$ is lowering operator for atom $i$ with ground state $|1 \rangle_i$ and the excited state $|2 \rangle_i$. The energy difference between both levels equals $\hbar \omega_0$ while $\omega_k$ is the frequency of a photon with  wave vector ${\bf k}$. Moreover, $a_{{\bf k}\lambda}$ is the annihilation operator for a photon with ${\bf k}$ and polarisation $\lambda$. The coupling strength between the atomic dipole moments and the photon mode $({\bf k}, \lambda)$ is given by the coupling constant $g_{{\bf k}\lambda}$. 

The final term in Eq.~(\ref{21}) is the laser Hamiltonian. Its role is to re-excite the particles after each photon emission. The first two terms in Eq.~(\ref{21}) describe the interaction between the atoms and the free radiation field. Whenever there is some population in the excited states $|2 \rangle_i$, energy can be transferred into the photon modes $({\bf k}, \lambda)$. The result is the dissipation of energy from the atoms into the surrounding field modes. In other words, the Hamiltonian (\ref{21}) entangles the state of the atoms with the free radiation field.

\begin{figure}[b]
\centerline{\epsfig{file=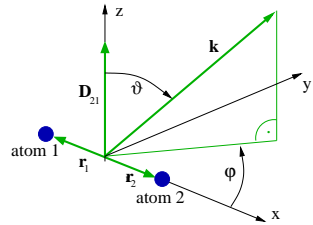,width=40mm}}
\caption{Coordinate system with the spatial angles $\vartheta$ and $\varphi$ characterising the direction of the wave vector ${\bf k}$. Here we assume that the atomic dipole moment ${\bf D}$ is perpendicular to the line connecting both atoms.}\label{doubleslit2}
\end{figure}

However, the setup shown in Figure \ref{doubleslit} cannot be described by the continuous solution of a Schr\"odinger equation. To take the possibility of spontaneous photon emissions into account, we also have to consider the environment. The experimental observation of radiating atoms suggests to model the environment by assuming rapidly repated measurements whether a photon has been emitted or not \cite{hegerfeldt}. In case of a click, the direction $\hat{\bf k}$ of the emitted photon is registered on the screen. In Ref.~\cite{Schoen} we showed with the help of Eq.~(\ref{21}) that the state of two atoms prepared in $|\psi \rangle$ equals, up to normalisation,
\begin{equation}
|\psi_{\hat{\bf k}} \rangle \equiv R_{\hat{\bf k}} \, |\psi \rangle
\end{equation}
with
\begin{equation} \label{r}
R_{\hat{\bf k}}  = \left( {3 A \over 8 \pi} \right)^{1/2} \, \sin \vartheta \, 
\left( \, {\rm e}^{-{\rm i} {\bf k}_0 {\bf r}_1} \, S_1^- +  {\rm e}^{-{\rm i} {\bf k}_0 {\bf r}_2} \, S_2^- \, \right) 
\end{equation}
and ${\bf k}_0 = (2 \pi \omega_0/c) \, \hat {\bf k}$ immediately after the emission of a photon in the $\hat {\bf k}$-direction. In the derivation of Eq.~(\ref{r}), it was assumed that the atoms both have a dipole moment ${\bf D}$ orthogonal to the line connecting them, as shown in Figure \ref{doubleslit2}. Moreover, ${\bf r}_1$, ${\bf r}_2$ and $A$ denote the positions and the spontaneous decay rate of the particles.  

\begin{figure}[b]
\centerline{\epsfig{file=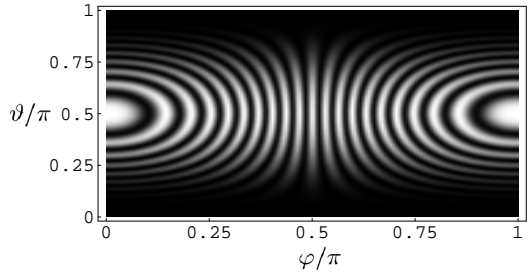,width=70mm}}
\caption{Density plot of the emission rate $I_{\hat{\bf k}}(\rho)$ for two continuously driven two-level atoms, $r = 10 \, \lambda_0$ and $\Omega = 0.3 \, A$. White areas correspond to spatial angles with maximal intensity and the definition of the angles $\varphi$ and $\vartheta$ is as shown in Figure \ref{doubleslit2}.}\label{doubleslit3}
\end{figure}

Eq.~(\ref{r}) is the key ingredient for the analysis of the interference pattern in the two-atom double-slit experiment \cite{Eichmann}. Let us assume for the moment that the atoms are again and again prepared in the same initial state $|\psi \rangle$. We now ask the question, what is the probability density to observe a ``click'' in a certain direction $\hat{\bf k}$. This probability density $I_{\hat{\bf k}}(\psi) $ is given by the norm squared of the state in Eq.~(\ref{r}),
\begin{equation} \label{i}
I_{\hat{\bf k}}(\psi) = \| \, R_{\hat{\bf k}} \, |\psi\rangle \, \|^2 \, .
\end{equation}
Since the reset operator (\ref{r}) is the sum of the reset operators of the cases, where there is either only atom 1 or only atom 2,
\begin{equation}
R_{\hat{\bf k}} = R_{\hat{\bf k}}^{(1)} + R_{\hat{\bf k}}^{(2)} \, ,
\end{equation}
the probability density (\ref{i}) is of the form
\begin{eqnarray} \label{int}
I_{\hat{\bf k}}(\psi) &=& \| \, R_{\hat{\bf k}}^{(1)}  \, |\psi \rangle 
+ R_{\hat{\bf k}}^{(2)}  \, |\psi\rangle \, \|^2  \nonumber \\
&=& \| \, R_{\hat{\bf k}}^{(1)} \, |\psi\rangle \, \|^2 + \| \, R_{\hat{\bf k}}^{(2)} \, |\psi\rangle \, \|^2 
+ \langle \psi | \,  R_{\hat{\bf k}}^{(1)\dagger} R_{\hat{\bf k}}^{(2)} + R_{\hat{\bf k}}^{(2)\dagger} R_{\hat{\bf k}}^{(1)} \, |\psi \rangle \, . \nonumber \\ 
\end{eqnarray}
This equation shows that the intensity of the light emitted from two atoms is not the same as the sum of the light intensities from two independent atoms (first two terms in Eq.~(\ref{int})). The difference is the {\em interference} term (third term in Eq.~(\ref{int})) which causes the photon emission in some directions to be more likely than the emission into others. If one replaces the pure state $|\psi \rangle$ by the stationary state $\rho$ of the atoms in the presence of continuous laser excitation, Eq.~(\ref{int}) can be used to calculate the interference pattern for the experimental setup in Figure \ref{doubleslit} \cite{Schoen}. The result is shown in Figure \ref{doubleslit3} and agrees very well with the observation in the experiment by Eichmann {\em et al.} \cite{Eichmann}. 

The origin of the spatial modulations in the interference pattern of the atoms is the wave behaviour of the excitation in the photon modes $({\bf k}, \lambda)$ prior to the detection of the photons. A photon does not really exist until it is actually observed on the screen. The laser field leads to a continuous re-excitation of the particles. The coupling between the atoms and the free radiation field then results in the transfer of energy into the free radiation field. Quantum mechanics tells us that it is only possible to detect an integer number of photons at any given point on the screen (c.f.~Figure \ref{doubleslit}). However, prior to the detection there can be more or less than one photon in each mode of the free radiation field.

Each detected photon is in general created by both atoms. Moreover, each photon leaves a trace in and contains information about all its respective sources. This is, of course, well known. Heisenberg wrote already in 1930, {\em It is very difficult  for us to conceive the fact that the theory of photons does not conflict with the requirements of the Maxwell equations. There have been attempts to avoid the contradiction by finding solutions of the latter which represent `needle' radiation (unidirectional beams), but the results could not be satisfactorily interpreted until the principles of the quantum theory had been elucidated. These show us that whenever an experiment is capable of furnishing information regarding the direction of emission of a photon, its results are precisely those which would be predicted from a solution of the Maxwell equations of the needle type (...)} \cite{Heisenberg}.

\section{Repeat-until-success quantum computing} \label{main}

\begin{figure}[b]
\centerline{\epsfig{file=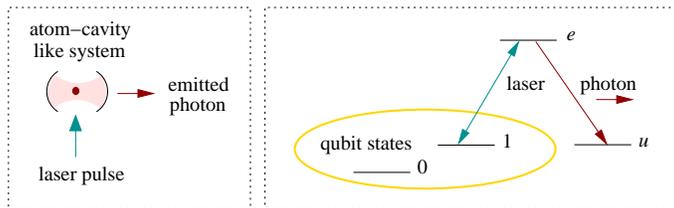,width=90mm}}
\caption{Experimental setup and atomic level configuration for the generation of a single photon on demand.}\label{pistol}
\end{figure}

If the photons emitted from the atoms do not exist until their detection and each detected photon leaves a trace in and contains information about all its respective sources, then photon emission should be a very useful tool to process information between distant stationary qubits. To show that this is indeed the case we now describe a quantum computing scheme based on this idea. The considered setup consists of a network of stationary qubits which can be used to generate single photons on demand. In such a setup, read-out measurements and single qubit rotations can be performed using laser pulses and standard quantum optics techniques as employed in recent ion trap experiments \cite{blatt,wineland}. 

More concretely, we now consider an atom-cavity like single photon source as shown in Figure \ref{pistol}. It consists of a single atom with a $\Lambda$-like level configuration trapped inside a resonant optical cavity. Alternatively, the atom can be replaced by an ``artificial atom'' like a quantum dot or a defect centre in a solid. To generate a single photon on demand, the atom should initially be prepared in $|1 \rangle$ and a laser pulse with a relatively slowly increasing Rabi frequency should be applied. Such a pulse transfers the atom into $|u \rangle$ and places exactly one photon into the cavity \cite{Law,Law2,Kuhn2,Mckeever}. From there it leaks out through the outcoupling mirror. Repumping the atom into its initial state results in the overall operation
\begin{equation} \label{one}
|1 \rangle ~~ \longrightarrow ~~  |1 ; 1_{\rm ph} \rangle \, .   
\end{equation}
The role of the cavity is to fix the direction of the spontaneously emitted photon so that it can be easily processed further.

Suppose each atom within a large network of single photon sources contains one qubit consisting of the two ground states $|0 \rangle$ and $|1 \rangle$ (c.f.~Figure \ref{pistol}). Then it is possible to generate a single photon on demand such that 
\begin{equation} \label{two}
\alpha \, |0 \rangle + \beta \, |1 \rangle ~~ \longrightarrow ~~  
\alpha \, |0;{\sf E} \rangle + \beta \, |1;{\sf L} \rangle \, . 
\end{equation} 
Here $|{\sf E} \rangle$ denotes an early and $|{\sf L} \rangle$ denotes a late photon. One way to implement the encoding step (\ref{two}) is to first swap the atomic states $|0 \rangle$ and $|1 \rangle$. Then a laser pulse with increasing Rabi frequency should be applied to perform the operation (\ref{one}). Afterwards, the states $|0 \rangle$ and $|1 \rangle$ should be swapped back and the photon generation process (\ref{one}) should be repeated at a later time. In the final state (\ref{two}), the atom is {\em entangled} with the state of the newly generated photon. The qubit is now double encoded in the state of the source as well as in the state of the photon. 

\begin{figure}[b]
\centerline{\epsfig{file=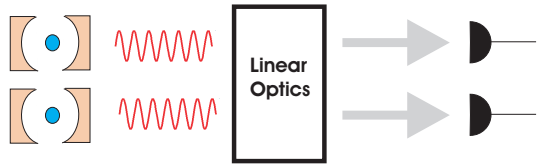,width=70mm}}
\caption{The experimental realisation of a universal two-qubit gate requires the generation of a photon within each of the sources involved. The two photons then pass within their coherence time through a linear optics network, which performs a photon pair measurement on them in a carefully chosen basis.}\label{pistol2}
\end{figure}

The encoding (\ref{two}) is the main building block for the realisation of an eventually deterministic entangling gate operation between two qubits. It requires the simultaneous generation of a photon in each of the involved single photon sources. Afterwards, the photons should pass, within their coherence time, through a linear optics network which performs a photon pair measurement on them (c.f.~Figure \ref{pistol2}). Suppose, the two qubits involved in the gate operation are initially prepared in the arbitrary two-qubit state
\begin{equation} \label{in}
|\psi_{\rm in} \rangle = \alpha \, |00 \rangle + \beta \, |01 \rangle + \gamma \, |10 \rangle + \delta \, |11 \rangle \, .
\end{equation}
Using Eq.~(\ref{two}), we see that the state of the system equals 
\begin{equation} \label{enc}
|\psi_{\rm enc} \rangle = \alpha \, |00;{\sf EE} \rangle + \beta \, |01;{\sf EL} \rangle + \gamma \, |10;{\sf LE} \rangle + \delta \, |11;{\sf LL} \rangle 
\end{equation} 
after the creation of the two photons. If the detectors indicate a photon pair measurement of a state of the form 
\begin{equation} \label{max}
|\Phi \rangle = |{\sf EE} \rangle + {\rm e}^{{\rm i} \varphi_1} \, |{\sf EL} \rangle 
+ {\rm e}^{{\rm i} \varphi_2} \, |{\sf LE} \rangle + {\rm e}^{{\rm i} \varphi_3} \, |{\sf LL} \rangle \, ,
\end{equation}
then the final state of the photon sources equals
\begin{equation}
|\psi_{\rm fin} \rangle  = \alpha \, |00 \rangle + {\rm e}^{-{\rm i} \varphi_1} \, \beta \, |01 \rangle 
+ {\rm e}^{-{\rm i} \varphi_2} \, \gamma \, |10 \rangle + {\rm e}^{-{\rm i} \varphi_3} \, \delta \, |11 \rangle \, .
\end{equation}
This state differs from the one in Eq.~(\ref{in}) by a unitary operation, namely a two-qubit phase gate. As we see below, if the state (\ref{max}) is a maximally entangled state, then this phase gate is one with maximum entangling power. This means that the described process can transform a non-entangled product state into a maximally entangled one.

However, the above photon pair measurement can only be used to perform an eventually deterministic gate operation, if a complete set of basis states can be found with each of them being of the form (\ref{max}). That such bases exist is well known. They are called mutually unbiased \cite{Wootters}, since their observation does not reveal any information about the computational states. Whatever state the photons are found in, the coefficients $\alpha$, $\beta$, $\gamma$ and $\delta$ remain unknown. Below we give an example of a mutually unbiased basis for two time bin encoded photons. The problem with linear optics is that it does not allow for complete Bell measurements \cite{Luetkenhaus,Luetkenhaus2}. At most two maximally entangled state can be distinguished. We therefore consider the basis states \cite{6}
\begin{equation}
|\Phi_{1,2} \rangle \equiv {\textstyle {1 \over \sqrt{2}}} \, \big( |{\sf x}_1 {\sf y}_2 \rangle \pm |{\sf y}_1 {\sf x}_2 \rangle \big) \, , ~~ |\Phi_3 \rangle \equiv |{\sf x}_1 {\sf x}_2 \rangle \, , ~~ 
|\Phi_4 \rangle \equiv |{\sf y}_1 {\sf y}_2 \rangle 
\end{equation}
with
\begin{eqnarray}  \label{xy}
&& |{\sf x}_1 \rangle =  {\textstyle {1 \over \sqrt{2}}} \, \big( |{\sf E} \rangle + |{\sf L} \rangle \big) \, , ~~
|{\sf y}_1 \rangle =  {\textstyle {1 \over \sqrt{2}}} \, \big( |{\sf E} \rangle - |{\sf L} \rangle \big) \, , \nonumber \\
&& |{\sf x}_2 \rangle =  {\textstyle {1 \over \sqrt{2}}} \, \big( |{\sf E} \rangle + |{\sf L} \rangle \big) \, , ~~ |{\sf y}_2 \rangle  =  {\textstyle {1 \over \sqrt{2}}} \, {\rm i} \,  \big( |{\sf E} \rangle - |{\sf L} \rangle \big) \, .
\end{eqnarray}
This definition implies 
\begin{eqnarray} \label{Phi}
&& |\Phi_{1,2} \rangle = \pm {\textstyle {1 \over 2}} \, {\rm e}^{\pm{\rm i} \pi/4}  \, \big( |{\sf EE} \rangle \mp {\rm i} \, |{\sf EL} \rangle \pm {\rm i} \,  |{\sf LE} \rangle - |{\sf LL} \rangle \big) \, , ~~ \nonumber \\
&& |\Phi_3 \rangle =  {\textstyle {1 \over 2}} \, \big( |{\sf EE} \rangle + |{\sf EL} \rangle + |{\sf LE} \rangle + |{\sf LL} \rangle \big) \, , \nonumber \\
&& |\Phi_4 \rangle =  {\textstyle {1 \over 2}} \, {\rm i} \, \big( |{\sf EE} \rangle -  |{\sf EL} \rangle - |{\sf LE} \rangle + |{\sf LL} \rangle \big) \, .
\end{eqnarray}
A comparison with Eq.~(\ref{max}) shows that the $|\Phi_i \rangle$ are indeed mutually unbiased. To find out which quantum gate operation belongs to which measurement outcome, we write the encoded state (\ref{enc}) as
\begin{equation}
|\psi_{\rm enc} \rangle = {\textstyle {1 \over 2}} \, \sum_{i=1}^4 |\psi_i \rangle |\Phi_i \rangle
\end{equation}
with
\begin{eqnarray}
|\psi_1 \rangle &=& {\rm e}^{-{\rm i} \pi/4} \, Z_1\big({\textstyle {1 \over 2}} \pi \big) \, Z_2 \big(-{\textstyle {1 \over 2}} \pi \big) \, U_{CZ} \, |\psi_{\rm in} \rangle \, , \nonumber \\
|\psi_2 \rangle &=& -{\rm e}^{{\rm i} \pi/4} \, Z_1\big(-{\textstyle {1 \over 2}} \pi \big) \, Z_2 \big({\textstyle {1 \over 2}} \pi \big) \, U_{CZ} \, |\psi_{\rm in} \rangle \, , \nonumber \\ 
|\psi_3 \rangle &=& |\psi_{\rm in} \rangle \, ,  \nonumber \\ 
|\psi_4 \rangle &=& -  {\rm i} \,  Z_1(\pi) \, Z_2(\pi) \, |\psi_{\rm in} \rangle \, .
\end{eqnarray}
Here $Z_i(\varphi)$ describes a one-qubit phase gate that changes the phase of an atom if it is prepared in $|1 \rangle$,
\begin{equation}
Z_i(\varphi) = {\rm diag} \, (0, {\rm e}^{-{\rm i}\varphi}) \, .
\end{equation}
Moreover, $U_{\rm CZ}$ denotes the controlled two-qubit phase gate
\begin{equation} \label{UZ}
U_{\rm CZ} = {\rm diag} \, ( 1, 1, 1, -1 ) 
\end{equation}
which changes the state of two atoms when they are prepared in $|11 \rangle$. The above equations show that a measurement of $|\Phi_1 \rangle$ or $|\Phi_2 \rangle$ results in the completion of the universal phase gate (\ref{UZ}) up to local operations which can be easily performed on the atom. A measurement of $|\Phi_3 \rangle$ or $|\Phi_4 \rangle$ yields the initial qubits up to local operations. Since the quantum information stored in the system is not lost at any stage of the computation, the above described steps can be {\em repeated until success}. On average, the completion of one RUS quantum gate requires two repetitions.

\begin{figure}[b]
\centerline{\epsfig{file=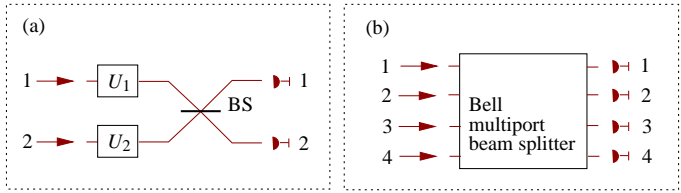,width=90mm}}
\caption{Two possible experimental setups for the realisation of the photon pair measurement in a mutually unbiased basis.}\label{repeat}
\end{figure}

Finally, we comment on the experimental realisation of the proposed quantum computing architecture. More details can be found in Refs.~\cite{6,9}. One possibility to realise the above described photon pair measurement is to convert the time bin encoding of the photonic qubits into a polarisation encoding. It is known that sending two polarisation encoded photons through a beam splitter results in a measurement of the states $|{\sf hv \pm  vh} \rangle$, $|{\sf hh} \rangle$ and $|{\sf vv}\rangle$. Measuring the states $|\Phi_i \rangle$ therefore only requires passing the photons through a beam splitter after applying the mapping 
\begin{equation}
U_i = |{\sf h} \rangle \langle {\sf x_i}| + |{\sf v} \rangle \langle {\sf y}_i| 
\end{equation}
on the photon coming from source $i$. The states $|x_i \rangle$ and $|y_i \rangle$ are given in Eq.~(\ref{xy}) (c.f.~Figure \ref{repeat}(a)). 

Alternatively, the photons could be send through a Bell multiport beam splitter, as shown in Figure \ref{repeat}(b). An early (late) photon from source 1 should enter input port 1 (3) and an early (late) photon from source 2 should enter input port 2 (4).  If $b_n^\dagger$ denotes the creation operator for a photon in output port $n$, one can show that the network transfers the basis states $|\Phi_i \rangle$ such that \cite{6}
\begin{eqnarray} \label{multiportresult}
|\Phi_1 \rangle & \to & {\textstyle {1 \over {\sqrt 2}}} \, \big(  b_1^\dagger b_4^\dagger - b_2^\dagger 
b_3^\dagger \big) \, |{\rm vac} \rangle \, , \nonumber  \\ 
|\Phi_2 \rangle & \to & - {\textstyle {1 \over {\sqrt 2}}} \, \big(  b_1^\dagger b_2^\dagger - b_3^\dagger 
b_4^\dagger  \big) \, |{\rm vac} \rangle \, , \nonumber  \\
|\Phi_3 \rangle & \to &  {\textstyle {1 \over 2}} \, \big( b_1^{\dagger \, 2}  - b_3^{\dagger \, 2} \big) \, |{\rm vac} \rangle \, , \nonumber \\ 
|\Phi_4 \rangle & \to & - {\textstyle {1 \over 2}} \, \big( b_2^{\dagger \, 2}  - b_4^{\dagger \, 2} \big) \, |{\rm vac} \rangle \, . 
\end{eqnarray}
Here $|{\rm vac} \rangle$ is the vacuum state with no photons in the setup. Eq.~(\ref{multiportresult}) shows that detecting two photons in the same output port indicates a measurement of the state $|\Phi_3 \rangle$ and $|\Phi_4 \rangle$, respectively. Finding the two photons in different output ports indicates a measurement of  $|\Phi_1 \rangle$ or $|\Phi_2 \rangle$.

When we use photon detectors with finite efficiencies and when the photon generation is not ideal, a failure of the two-qubit gate operation does not always leave the qubits undisturbed. Consequently, the RUS procedure fails occasionally. However, RUS quantum gates can still be used for quantum computing. As recently shown by Barrett and Kok \cite{5}, it is possible to use entangling operations with arbitrarily high photon losses and finite success rates to efficiently generate graph states for one-way quantum computing \cite{10}. Combining the loss-tolerant mechanism described in Ref.~\cite{5} with the RUS quantum gate \cite{6} leads to a quantum computer architecture that is scalable and robust against inevitable losses but, most importantly, does not require the coherent control of qubit-qubit interactions.

\section{Conclusions} \label{conc}

In the first part of this manuscript, we discussed a recent two-atoms double-slit experiment \cite{Scully82,Eichmann}. This experiment showed that spontaneously emitted photons carry the information about and are entangled with all their respective sources (see also Refs.~\cite{Monroe,weber}). In the second part, we used the interference of spontaneously emitted photons within a linear optics setup to perform entangling gate operations between distant qubits. More concretely, we described the idea of Repeat-Until-Success (RUS) quantum computing \cite{6}. RUS quantum gates do not require the use of apriori created entanglement nor does it require to feed a photon back into a photon source. It also does it require the coherent control of a direct qubit-qubit interaction. We are therefore optimistic that RUS quantum computing opens new perspectives for finding really scalable and robust quantum computing architectures. 

\section*{Acknowledgments}

A. B. acknowledges support from the Royal Society and the GCHQ. This work was supported in part by the EU and the UK EPSRC through the QIP IRC.

\end{document}